\documentclass[10pt,twocolumn,letterpaper]{article}

\usepackage{cvpr}
\usepackage{times}
\usepackage{epsfig}
\usepackage{graphicx}
\usepackage{amsmath}
\usepackage{amssymb}

\usepackage[pagebackref=true,breaklinks=true,letterpaper=true,colorlinks,bookmarks=false]{hyperref}

\cvprfinalcopy 

\begin{document}

\title{Microstructure Surface Reconstruction from SEM Images:\\An Alternative to Digital Image Correlation (DIC)}

\author{Khalid El-Awady\\
Department of Computer Science\\
Stanford University\\
{\tt\small kae@stanford.edu}
}

\maketitle

\begin{abstract}
   We reconstruct a 3D model of the surface of a material undergoing fatigue testing and experiencing cracking. Specifically we reconstruct the surface depth (out of plane intrusions and extrusions) and lateral (in-plane) motion from multiple views of the sample at the end of the experiment, combined with a reverse optical flow propagation backwards in time that utilizes interim single view images. These measurements can be mapped to a material strain tensor which helps to understand material life and predict failure. This approach offers an alternative to the commonly used Digital Image Correlation (DIC) technique which relies on tracking a speckle pattern applied to the material surface. DIC only produces in-plane (2D) measurements whereas our approach is 3D and non-invasive (requires no pattern being applied to the material).
\end{abstract}


\section{Introduction}
We study 3D micsrostructure surface modeling over time for a material undergoing fatigue testing. Fatigue refers to the initiation and propagation of cracks in a material due to cyclic loading. The experimental study of fatigue at the microstructure level is often done through visual inspection of test materials using scanning electron microscopes (SEMs). 

The scanning electron microscope (SEM) is the most commonly used tool for imaging of microstructures . These are structures with dimensions around 1 micron ($10^{-6}$m). The image formed is a projected one, similar to traditional optical cameras. 

\begin{figure}[t]
\begin{center}
    \includegraphics[width=1\linewidth]{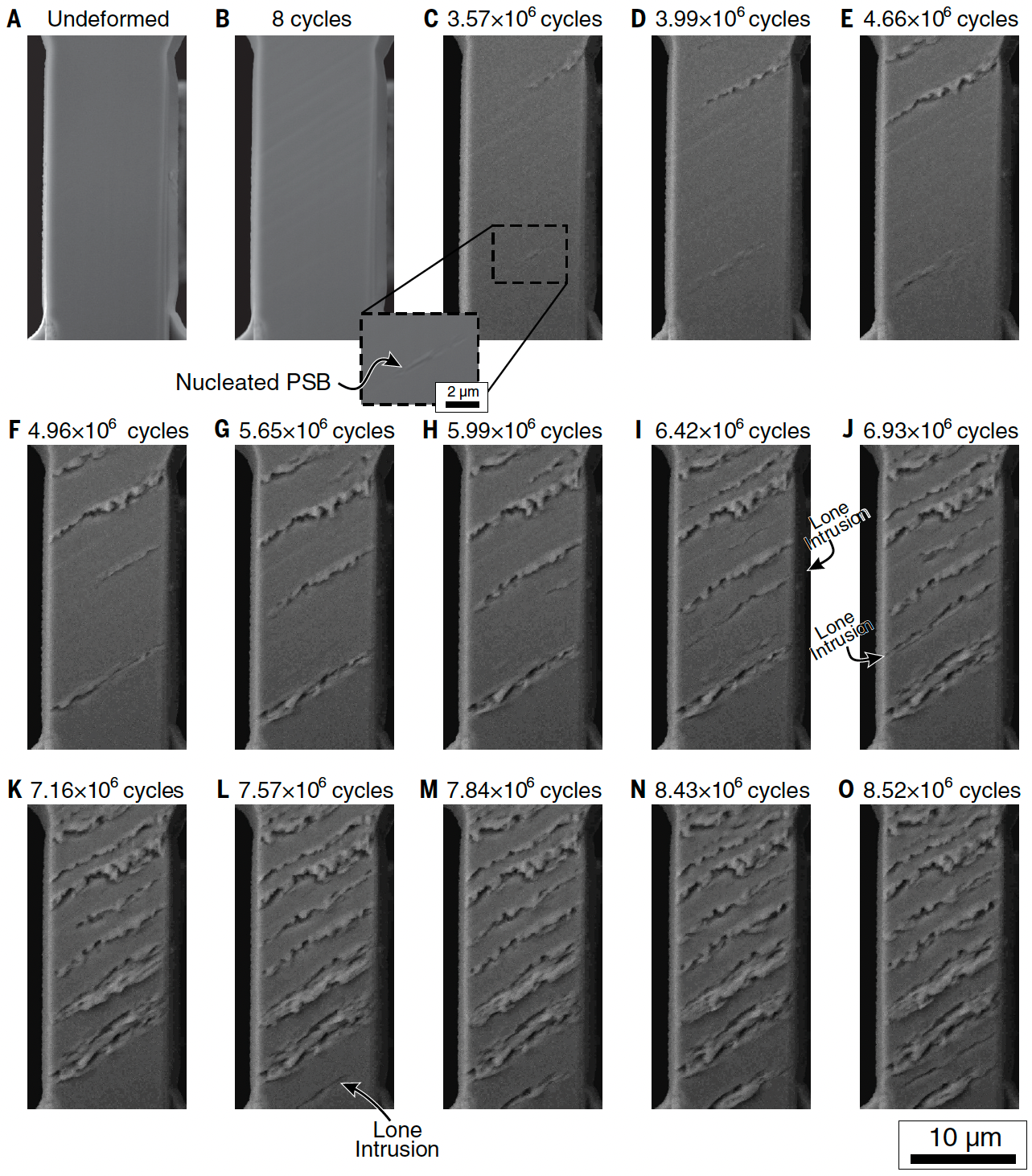}
\end{center}
   \caption{Fatigue experiment conducted on a material with successive SEM images generated at progressive parts of the experiment showing the formation and propagation of cracks (image from \cite{Lavenstein}).}
\label{fig:science}
\end{figure}

Figure (\ref{fig:science}) from \cite{Lavenstein} shows an example of such a study. A small rectangular block of nickel (roughly 10 $\mu$m $\times$ 30 $\mu$m) is placed under the SEM and repeatedly subjected to cycles of tension and compression in the lengthwise direction ($y$ direction in the image). Over time cracks begin to form at internal planes at certain known orientations that relate to the crystal structure of the material. 

\begin{figure}[h]
\begin{center}
    \includegraphics[width=01\linewidth]{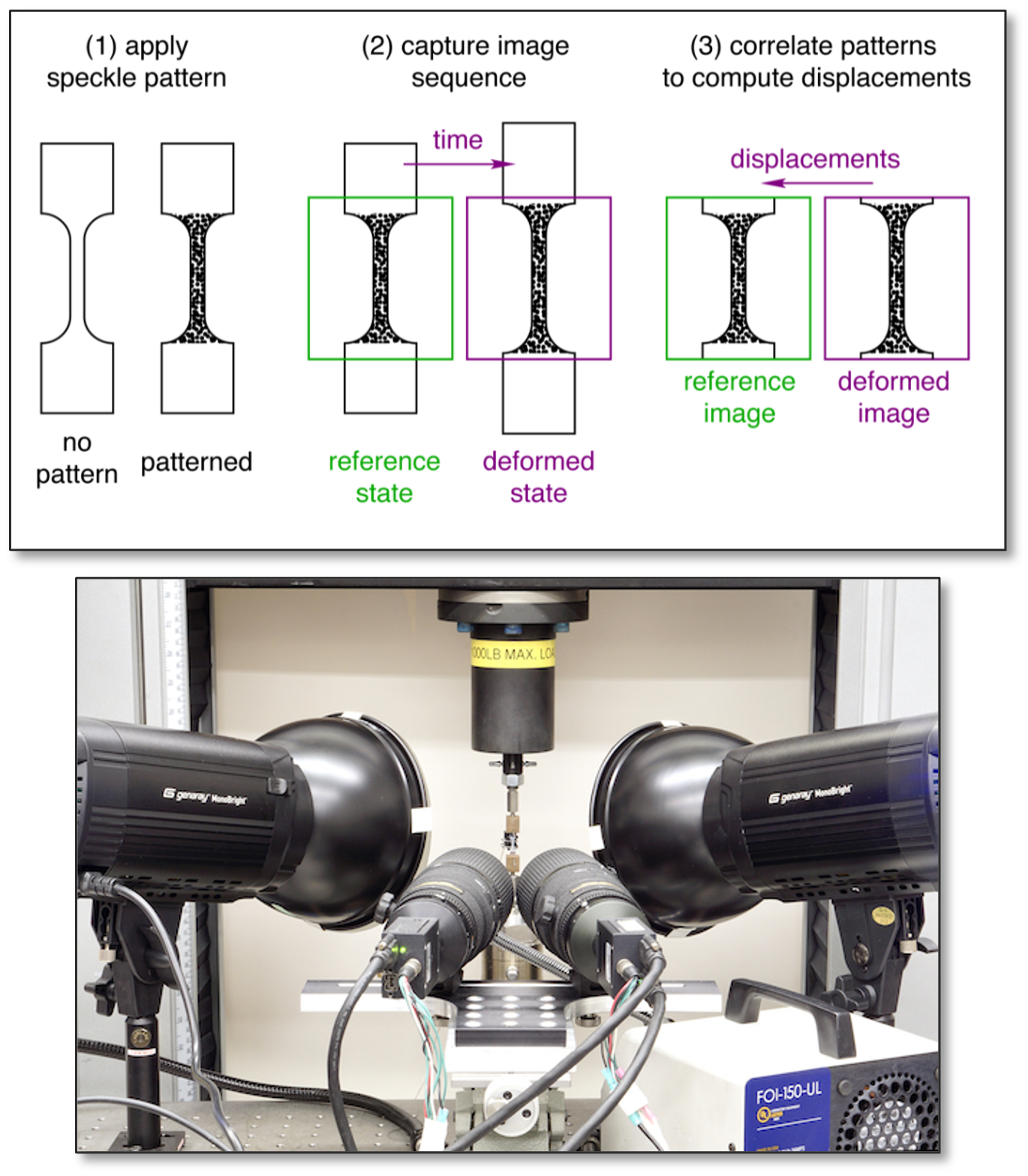}
\end{center}
\caption{Digital Image Correlation setup. Cameras are used to record changes in a 'speckle' pattern that is imprinted on the sample being tested. (Images taken from https://digitalimagecorrelation.org/)}
\label{fig:DIC}
\end{figure}

The images show the growth of these cracks in both number and size. It is of interest to material scientists to understand the deformation of the surface which can be mapped to a strain tensor. Material Scientists then use the strain tensor to predict material life and failure modes. One technique commonly used today is Digital Image Correlation \cite{Yang2010}. In this technique, shown in figure (\ref{fig:DIC}), a known so-called 'speckle' pattern is imprinted on the sample being tested and images are correlated over time to estimate how the surface is deforming. This technique, though, is invasive and limited to 2D estimates of the surface deformation. 

We present an alternative approach that estimates the magnitude and shape of the extrusions and intrusions on the surface -- in other words the depths and heights of the peaks and valleys relative to the surface, in addition to the lateral (in-plane) deformation. The surface extrusions/intrusions can more accurately inform the correlation between fatigue life and slip localization. Recent studies suggest that there is a direct correlation between those and that fatigue life can be predicted from these out-of-plane displacements. Our goal is to detect variations in the z-direction on the order of 0.1 $\mu$m.

During the experiment, the SEM cannot be re-oriented so all the images obtained are from a fixed head-on viewpoint. Once the experiment is complete, though, the SEM can be re-oriented to obtain additional views.

Our approach is summarized as follows. First we use stereo metrology to estimate the surface roughness at the end of the experiment using two images of the sample at the end of the experiment from different camera orientations. Then we use optical flow techniques applied to single images of the sample during testing to estimate the lateral (in-plane) evolution of the cracks. 

\section{Related Work}

3D imaging of microstructures is possible directly through other microscopy techniques, such as scanning tunneling microscopes (STMs) and atomic force microscopes (AFMs). The former, though, requires extremely thin materials (on the order of nm's), and the latter is limited to depths of $\sim5 \mu$m and is costly. Therefore there has evolved a relatively rich set of literature exploring the reconstruction of 3D models from 2D SEM images in recent years. These include the studies of \cite{Taftia, Kozikowski, Rehman} for example. These studies produce disparity maps from stereo metrology, but do not specifically study the geometry of cracks. 

Also, as mentioned earlier, the technique of Digital Image Correlation \cite{Yang2010}, is commonly used to study 2D surface deformation of samples undergoing fatigue testing. 

Our contribution (to our knowledge) is thus to offer an approach that combines stereo imaging from the end of the experiment, with non-invasive optical flow analysis applied to mono imaging during the experiment to provide a complete 3D view of the deformation of the sample undergoing fatigue.

\section{Dataset}
All the images were provided by Prof. J. El-Awady, Associate Professor of Mechanical Engineering at John Hopkins University, co-author of \cite{Lavenstein} (and my brother) \cite{Jelawady}. The images are all original acquisitions from his lab. 

\begin{figure}[h]
\centering
    \includegraphics[width=1\linewidth]{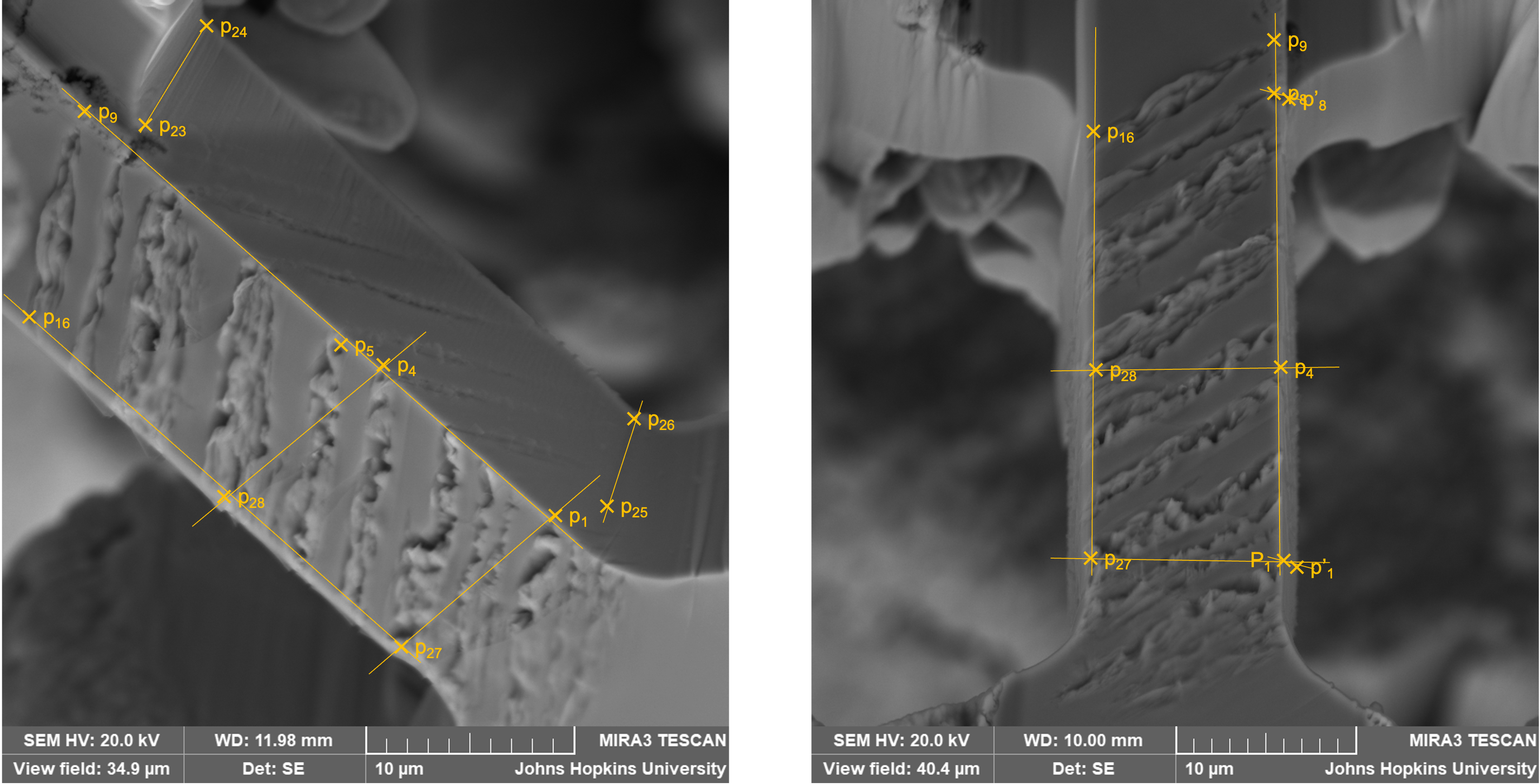}
\caption{Two views of the sample at the conclusion of the experiment with identified parallel lines from 3 normal planes.}
\label{fig:vanishing_points1}
\end{figure}

\section{End State Surface Roughness Estimation}

\subsection{Vanishing Point Analysis}
Figure (\ref{fig:vanishing_points1}) shows two views of the sample at the end of the experiment. We have overlaid in yellow a pair of parallel lines in three normal planes that are identified visually from the figures. We proceed to compute the vanishing points and from them extract the camera's intrinsic matrix and the rotation matrix between the cameras \cite{Hata2}:
\begin{eqnarray*}
    K & = & \left[ \begin{array}{ccc}
                    1704 & 0 & 1300 \\ 
                    0 & 1704 & -1604 \\ 
                    0 & 0 & 1
                    \end{array} \right] \\
\end{eqnarray*}

{\footnotesize
\texttt{Angle around z-axis (pointing out of camera): 46.1 degrees}

\texttt{Angle around y-axis (pointing vertically): 3.2 degrees}

\texttt{Angle around x-axis (pointing horizontally): -62.1 degrees}
}

The identified angles appear consistent with visual inspection. 

\subsection{Epipolar Analysis} \label{EpipolarSection}
\begin{figure}[h]
\centering
    \includegraphics[width=1\linewidth]{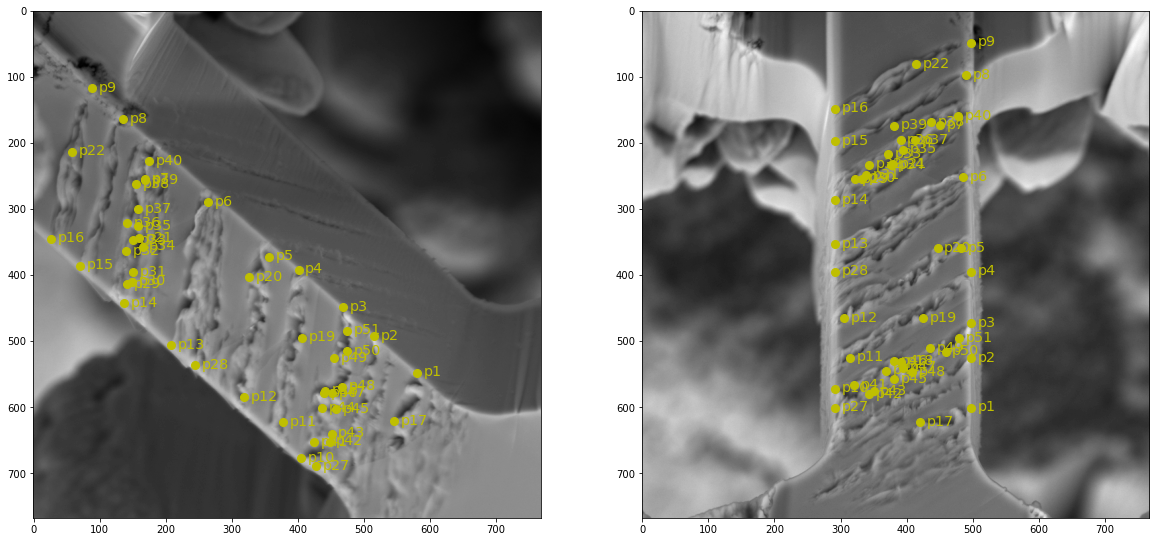}
\caption{Corresponding points identified visually for the two images at the end of the experiment.}
\label{fig:corresponding_points}
\end{figure}

\begin{figure}[h]
\centering
    \includegraphics[width=1\linewidth]{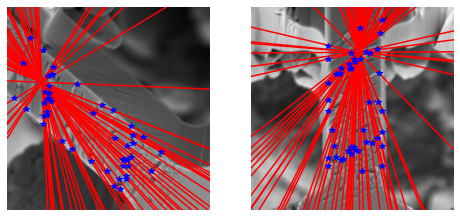}
\caption{Computed epipoles for the images using the normalized eight-point algorithm.}
\label{fig:epipoles_normalized}
\end{figure}

Next we apply epipolar analysis as presented in \cite{Hata3}. We begin by identifying corresponding points between the images as shown in figure (\ref{fig:corresponding_points}). The variations in the shape of the cracks provide a rich environment for identifying corresponding points. In total about 50 such points were identified across the image. The computed epipoles using these points are shown in figure (\ref{fig:epipoles_normalized}).

Note that experimentation showed that estimates of the epipoles was sensitive to the number of points used. Initial attempts to estimate the epipoles with less than 20 points resulted in significant variations in the estimated epipole. Once we increased corresponding points to 30 the estimate stabilized. 

\begin{figure}[h]
\centering
    \includegraphics[width=1\linewidth]{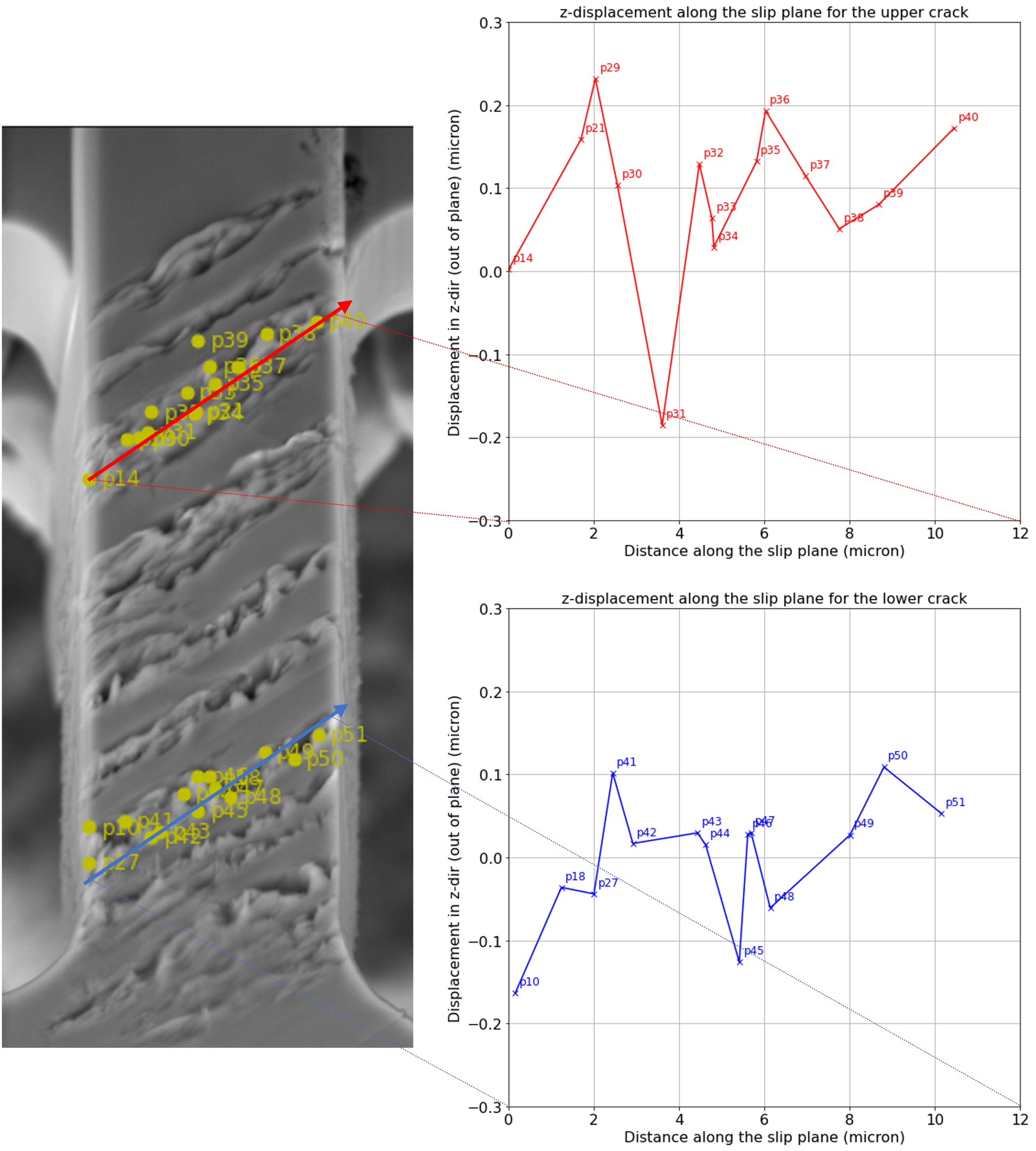}
\caption{Z-deformation for two of the cracks.}
\label{fig:two_cracks}
\end{figure}

Using these results we can apply the Tomasi and Kanaede Factorization Method to determine the 3D structure of the scene. The resultant estimates of depths are shown in the following figures. In figure (\ref{fig:two_cracks}) we show the estimated heights / depths relative to the surface along two cracks. Note the horizontal direction in the graphs represents the distance along the crack which is at an approximately $35^\circ$ angle to the horizontal image in the image. 

\begin{figure}[h]
\centering
    \includegraphics[width=1.1\linewidth]{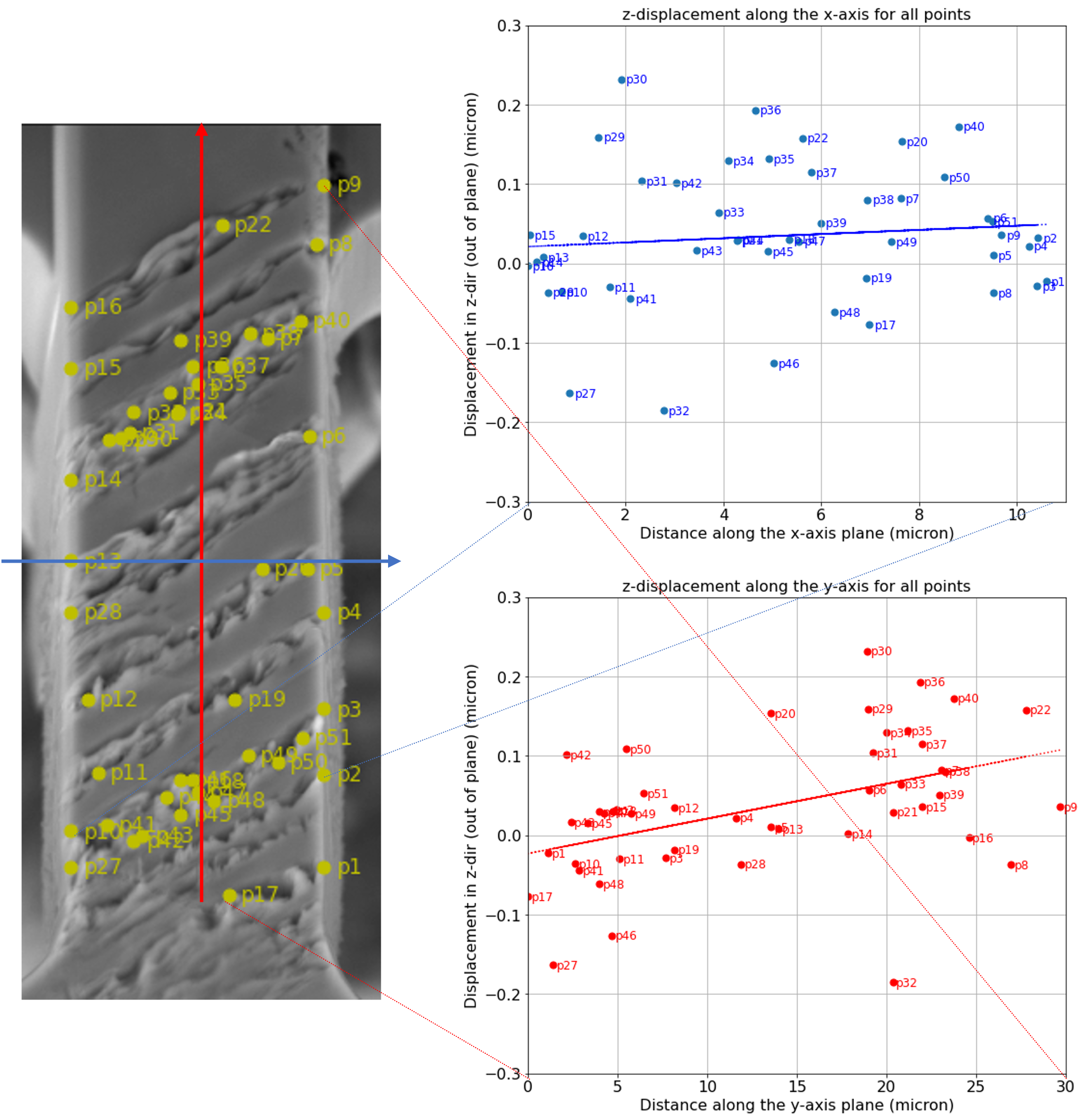}
\caption{Z-deformation for two of the cracks.}
\label{fig:all_points_projections2}
\end{figure}

Now consider all the points. In figure (\ref{fig:all_points_projections2}) we show the projection of the points along both the image x and y axes (i.e., the x-z plane projection and the y-z plane projection). This helps to show the range of z-displacements across the points. We see that the range of displacement is approximately $\pm 0.2 \mu m$. We also notice a slight trend of more positive displacement (protrusion) near the top of the image (high y-value) and more indentation at the bottom. There is a very modest trend of more protrusion to the right than the left. 

\subsection{Result Validation}
Review of these results with Prof. J. El-Awady indicate the results are very reasonable. The sample that was studied in the images was nickel, and while Prof. El-Awady does not have a direct measurement that can be used to independently to verify our estimates, there is literature where measurements have been done on other materials. \cite{Man2009, Polak2009, Polak2015}. These show intrusions and extrusions of the order of magnitude of 0.02 -- 0.3 micron in stainless steel, and as high 0.75 micron in Copper. Professor El-Awady's writes: "the conclusion is that an extrusion of 200 nm is very reasonable according to existing literature so far."

Prof. El-Awady further confirmed that the result of figure (\ref{fig:all_points_projections2}) showing a correlation between vertical position and the z-deformation makes sense from a physical standpoint that relates to the machining of the sample that results in increased strength at the lower part of the sample. 

While these comments are anecdotal, we take these are encouragement that our method produces plausible results and could be further calibrated with some direct measurements in the future. 

\section{Optical Flow and Lateral Movement}
Now that we have an estimate of the surface roughness in the form of point depth estimates derived from stereo views of the final sample state, we consider the evolution of these points from the original undeformed state. This will provide insight into the lateral (x-y) movement of the points over time. Recall that while the experiment is being conducted we only have access to a single fixed front-view image of the sample.

We use the Lucas-Kanade Opticalflow method to estimate the lateral movement associated with the microcracks. We employ it using two separate sets of initial features to track. The first uses features identified using the Tomasi-Shi corner detection approach. While we will not have depth estimates for these points, they nonetheless provide a baseline for comparison of estimated lateral movement. Then we will use our corresponding points with depth estimates from section \ref{EpipolarSection}.

\subsection{Lucas-Kanade Opticalflow with Tomasi-Shi Corner Features}

\begin{figure}[th]
\begin{center}
    \includegraphics[width=1\linewidth]{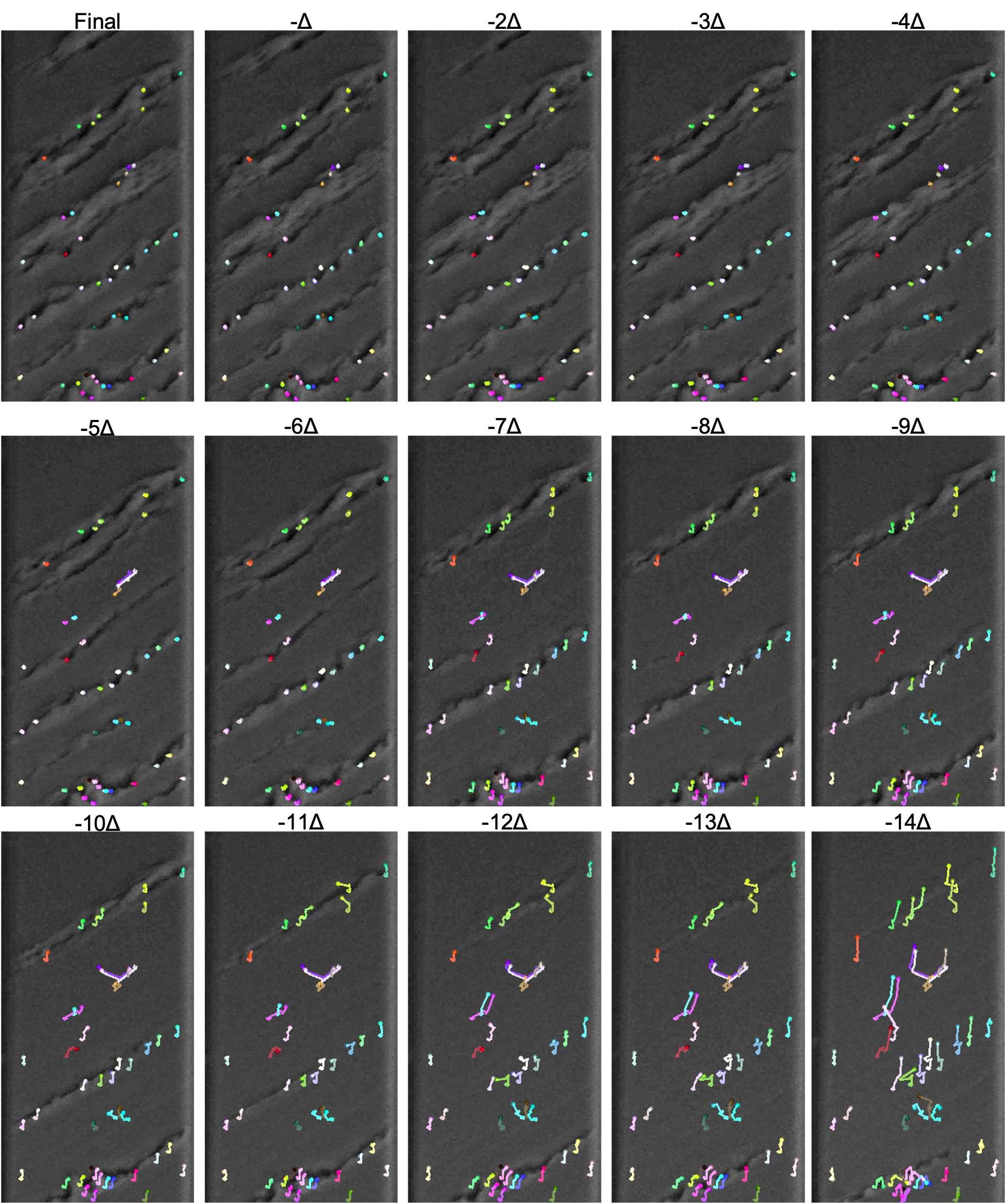}
\end{center}
   \caption{(Reverse) evolution of microcracks in the sample showing the movement of 45 points identified by the Tomasi-Shi feature detection algorithm. Each image represents average of 80 min time difference and 500,000 cycles.}
\label{fig:ShiTomasiRev}
\end{figure}

Figure (\ref{fig:ShiTomasiRev}) shows the optical flow estimated using the Lucas-Kanade method implemented in OpenCV \cite{OpenCV} using corner features identified using the Tomasi-Shi corner features. The figure shows a series of 15 sample images in \underline{reverse} chronological order with the top left image being the end state of the sample. The lower right image is earliest image of the sample where some amount of deformation is still visible. The sample are approximately 80 minutes and 500,000 cycles apart. 

We experiment with the parameters of the Tomasi-Shi algorithm to find a set of 45 points that persist across all images. We find this is best controlled by the 'quality level' parameter. We set it to 0.35, along with a minimum distance of 7 and block size of 7. The trajectories of the points is shown overlayed on the images. 

\begin{figure}[h]
\begin{center}
    \includegraphics[width=0.7\linewidth]{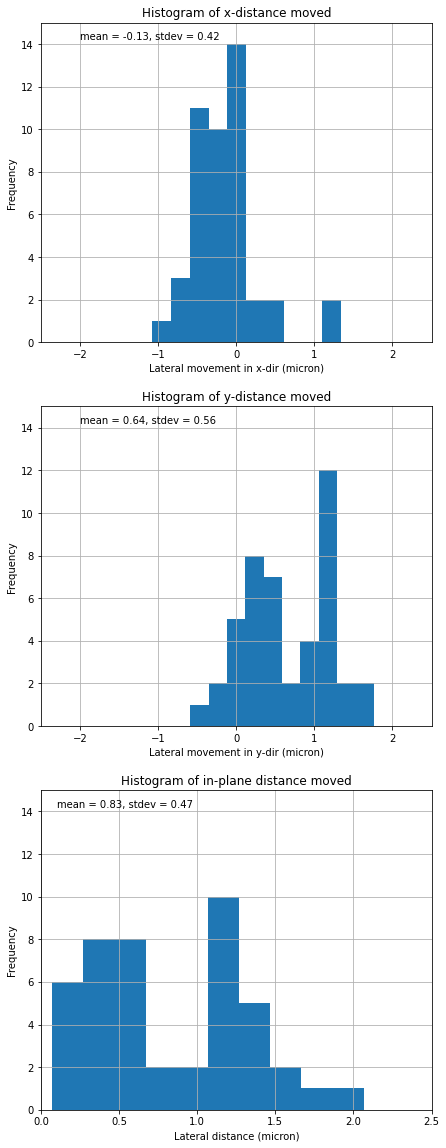}
\end{center}
   \caption{Histogram of the lateral movement of points tracked by the Lucas-Kanade algorithm using Tomasi-Shi detected corners.}
\label{fig:ShiTomasiHist}
\end{figure}

Figure (\ref{fig:ShiTomasiHist}) shows the histogram of these lateral movements from start to finish. We see that on average the points are moving to the left, the negative $x$ direction, by an average of 0.13 $\mu$m, and down, the positive $y$ direction, by an average 0.64 $\mu$m. The bottom plot shows the overall in-plane distance moved by the points, which averages 0.83 $\mu$m, though the distribution shows what appears to be a bi-modal distribution with about half the points moving around 1.5 $\mu$m and the other half only moving about 0.5 $\mu$m.

\subsection{Lucas-Kanade Opticalflow with Depth-Estimated Features}
\begin{figure}[h]
\begin{center}
    \includegraphics[width=1\linewidth]{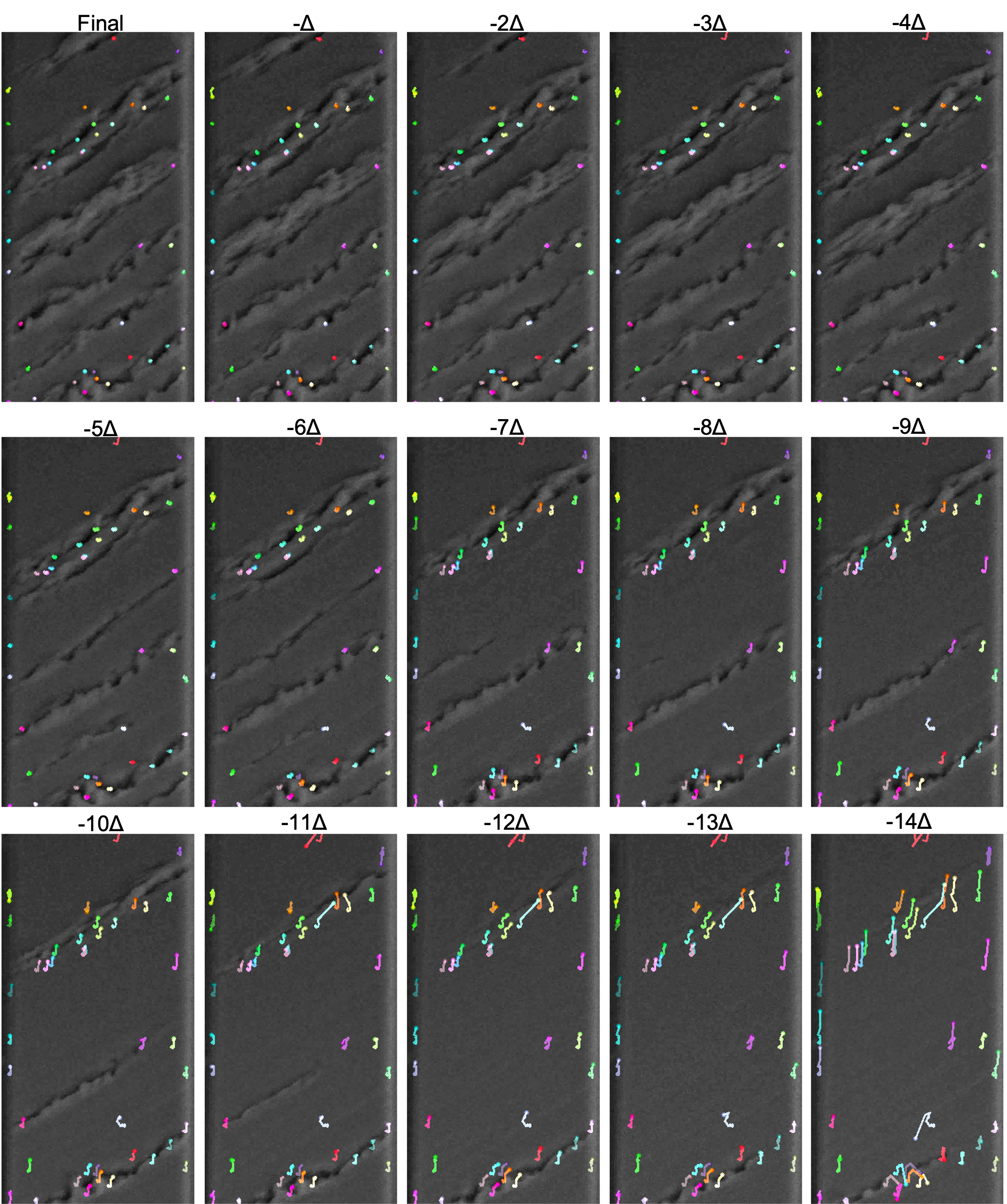}
\end{center}
   \caption{(Reverse) evolution of microcracks in the sample showing the movement of 41 points whose depth was estimated in section \ref{EpipolarSection}. Each image represents average of 80 min time difference and 500,000 cycles.}
\label{fig:DepthRev}
\end{figure}

\begin{figure}[h]
\begin{center}
    \includegraphics[width=0.8\linewidth]{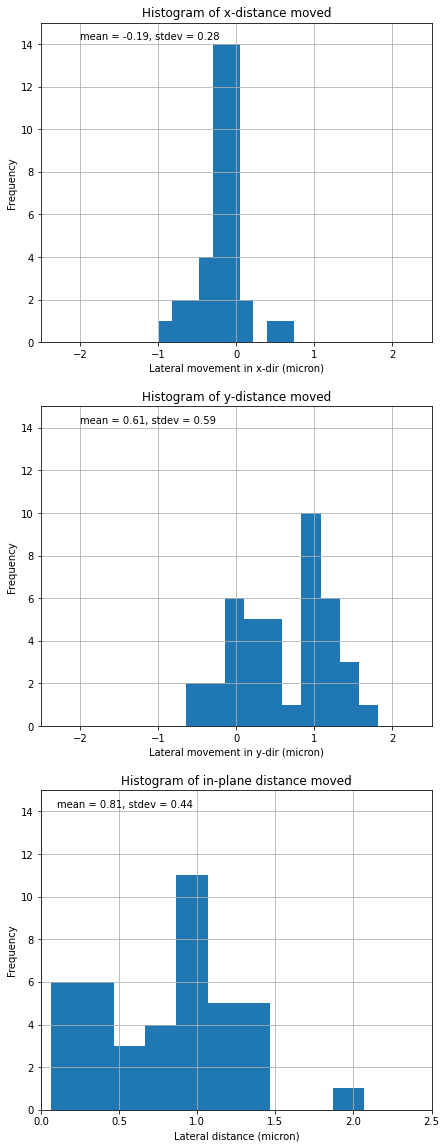}
\end{center}
   \caption{Histogram of the lateral movement of points whose depths were estimated in section \ref{EpipolarSection}.}
\label{fig:DepthHist}
\end{figure}

We now consider the Lucas-Kanade optical flow algorithm initialized using the set of points whose depths were estimated in section \ref{EpipolarSection}. Figure (\ref{fig:DepthRev}) shows the estimated tracks. The histogram associated with these lateral movements is shown in figure (\ref{fig:DepthHist}).

Comparing the results of figures (\ref{fig:ShiTomasiHist}) and (\ref{fig:DepthHist}) we see the shapes of the distributions and magnitudes and directions of the movements are consistent. This gives us a level of confidence in the use of these points. 

It is worth noting, though, that as the reverse optical flow approaches the earliest images where the deformation is small and most of the sample is crack free, the image exhibits broad areas of homogeneous pixel values and the estimated motion appears to make relatively large jumps. This is particularly noticeable in the last (earliest) frame. This is taken to be a limitation of the approach and not a physical outcome. 

\begin{figure}[h]
\begin{center}
    \includegraphics[width=0.7\linewidth]{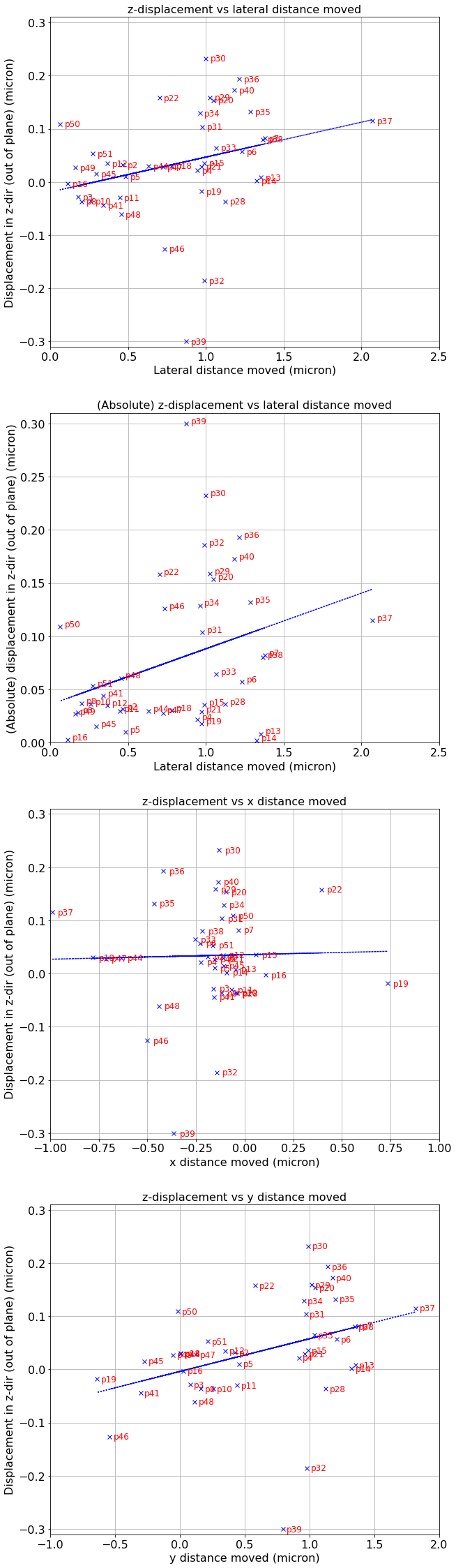}
\end{center}
   \caption{Plot of lateral (x-y) movement vs movement in the z-direction.}
\label{fig:zvslat}
\end{figure}

Now we investigate the correlation between the lateral and depth movement of the selected points. Figure (\ref{fig:zvslat}) shows a scatter plot of the estimated lateral vs depth movement. The top left shows the z vs. total distance moved, the top right is the absolute value of z vs distance moved, the lower left is z vs x, and lower right is z vs y. From the figures we notice a strong correlation between the magnitude of the movement in the y-direction and the magnitude of the final depth of the points. 

\subsection{Result Validation}
Discussions with Prof. J. El-Awady suggest that the lateral motion estimated from optical flow largely agrees with theoretical models. The cracks develop along certain slip planes within the material and these planes are purposely oriented so that all material motion happens in the y-z plane. This means that displacements in the x-direction are theoretically zero (or small) but movement in the y-direction expected. This conforms to the results presented above. The only exception to the expected results is the relatively large motions estimated for a few points in the final (earliest) frame due to the noted anomaly of the undeformed surface being largely of homogeneous pixel color.

\section{Conclusion and Future work}
We show how to reconstruct a 3D model of surface roughness from SEM images of a material undergoing fatigue testing. During the experiment, the camera (SEM) is fixed and produces only frontal views of the sample. At the conclusion of the experiment the sample can be detached and multiple views (at higher resolution) can be obtained. We use the views at the end of the experiment to reconstruct a 3D model of the surface roughness. Then, we use optical flow techniques to propagate the final surface back in time to obtain an estimate of the lateral movement of the points as well. Together the lateral and depth movement provide a comprehensive understanding of the surface deformation of the sample. 

Experts in the field have indicated that all our estimates of both out-of-plane deformation as well as in-plane lateral deformation are of a magnitude and shape consistent with theoretical expectation. Thus it represents a promising non-invasive alternative to the current state of the art, Digital Image Correlation, which requires imprinting a speckle pattern on the sample, and is limited to just 2D estimates unless it is augmented by further costly measurements such as x-ray tomography. 

A number of next steps are warranted:
\begin{itemize}
    \item The estimates should be confirmed experimentally. Tools such as atomic force microscopes can be used to obtain a measurement of the surface roughness. Samples can also be sectioned and examined to obtain depth measurements. While this is typically not done due to the cost and often limited access to such tools, the promise of the technique presented warrants performing this work. 
    
    \item Once the images are obtained, the most tedious aspect was manual selection of corresponding points. An investigation of automating this process should be undertaken. Tomasi-Shi corner features appears a promising approach as a starting point for feature identification. Our experiments suggest it will select prominent features along the cracks much like I did visually. We should investigate how this can be used as part of an automated process (or at least semi-automated) to identify the corresponding points. 
    
    \item Finally the surface deformation should be mapped to a strain tensor, which is the physical quantity that material scientists are ultimately seeking. 
\end{itemize}

\section{Acknowledgements}
The author would like to thank Andrey Kurenkov for his assistance and insightful suggestions about areas of investigation. Thank you, Andrey, for generously sharing your ideas and guidance. 

Further, all the data was courtesy of Prof. J. El-Awady of John Hopkins University. Prof. El-Awady provided the motivation for the work and also generously assessed, reviewed, commented on, and validated the relevance and quality of the results. 


{\small
\bibliographystyle{plain}

}

\end{document}